\documentclass[10pt,final,journal,twocolumn]{IEEEtran}
\ifCLASSINFOpdf

\else
\usepackage{algorithm}
\usepackage{booktabs}

\fi
\usepackage{cite}
\usepackage[cmex10]{amsmath}
\usepackage{amsthm}
\usepackage{algorithmic}
\usepackage{stfloats}
\usepackage{multicol,multienum}
\usepackage{multirow}
\usepackage{amssymb}
\usepackage{url}
\usepackage{amsmath}
\usepackage{xcolor}
\usepackage[dvips]{graphicx}
\usepackage{subfigure}

\newenvironment{shrinkeq}[1]
{ \bgroup
\addtolength\abovedisplayshortskip{#1}
\addtolength\abovedisplayskip{#1}
\addtolength\belowdisplayshortskip{#1}
\addtolength\belowdisplayskip{#1}}
{\egroup\ignorespacesafterend}

\begin{document}
\hyphenation{op-tical net-works semi-conduc-tor}


\title{Robust and Secure Beamforming for Intelligent Reflecting Surface Aided mmWave MISO Systems}
\setlength{\parskip}{0cm}
\author{Xingbo Lu, Weiwei Yang, Xinrong Guan, Qingqing Wu, Yueming Cai
\thanks{This work was supported by the Natural Science Foundations of China (No. 61671474 and No. 61671939).
X. Lu, W. Yang, X. Guan and Y. Cai are with Army Engineering University of PLA, Nanjing 210007, China. Q. Wu is with the Department of Electrical and Computer Engineering, National University of Singapore, Singapore 117583.
(Email: xblu1221@126.com, wwyang1981@163.com, geniusg2017@gmail.com, elewuqq@nus.edu.sg, caiym@vip.sina.com. \emph{Corresponding author: Weiwei Yang.})}
}
\IEEEpeerreviewmaketitle
\maketitle
\vspace{-0in}
\begin{abstract}
In this letter, we investigate the robust and secure beamforming (RSBF) in an intelligent reflecting surface (IRS) aided millimeter wave (mmWave) multiple input single output (MISO) system, where multiple single antenna eavesdroppers (Eves) are arbitrarily distributed nearby the legitimate receiver. Considering the channel state information (CSI) of Eves' channels is imperfectly known at the legitimate transmitter, the RSBF design problems to maximize the worst case of achievable secrecy rate (ASR) are formulated under the total transmission power and unit-modulus constraints. Since the problems are difficult to solve optimally due to their nonconvexity and coupled variables, we substitute the wiretap channels by a weighted combination of discrete samples and propose a RSBF scheme based on alternating optimization and semidefinite relaxation (SDR) techniques, for both colluding and noncolluding eavesdropping scenarios. Simulation results show that the proposed RSBF scheme can effectively improve the ASR and also outperforms other benchmark schemes.
\end{abstract}


\IEEEpeerreviewmaketitle
\vspace{-1mm}
\section{Introduction}
\vspace{-1mm}
Millimeter wave (mmWave) communication has been considered as a promising technology for next generation wireless communication systems, due to its abundant spectrum and thus unrivaled data rates \cite{Xiao2017Millimeter}. However, the substantial propagation loss renders the signals at mmWave band highly susceptible to be blocked by obstacles, thus resulting in a short coverage range. To address this issue, massive array antennas can be deployed, which, however, lead to unaffordable hardware cost and power consumption \cite{Yang2017Channel}. 
Recently, intelligent reflecting surface (IRS) has emerged as a promising technology to achieve high spectrum efficiency in a cost-effective manner [3]-[6]. Specifically, IRS is a software-controlled metasurface composed of a large number of reconfigurable and passive reflecting elements whose phase shift can be adaptively adjusted by a smart controller. By smartly adjusting the reflection coefficients, the reflected signals can be enhanced or weakened at different receivers. Due to the significant passive beamforming gain, IRS can be incorporated  into the mmWave communication systems to extend the coverage and reduce the need for radio frequency (RF) chains [5] \cite{Cao2019Intelligent}.

On the other hand, physical layer security (PLS) has been intensively investigated in recent years due to its capability of enhancing traditional secrecy system from the perspective of information theory. Since the IRS is able to configure the wireless channel in real-time via passive reflection, it has great potential in improving the secrecy rate [8]-[10]. Specifically,
an IRS-aided secure single user multiple input single output (MISO) system was investigated in \cite{Cui2019Secure}. By adaptively adjusting the IRS's phase shifts, the received signal at Bob is constructively strengthened but severely weakened with that at Eve, even though the legitimate and wiretap channels are highly correlated. An artificial-noise-aided secure wireless transmission scheme were proposed in the IRS-aided MIMO communication systems \cite{Hong2019Artificial}. The multiple Eves scenario was extended in \cite{Guan2019Intelligent}, in which the active beamforming (BF) vector with jamming and passive BF vector were jointly optimized to maximize the achievable secrecy rate (ASR). 

\begin{figure}
\setlength{\belowcaptionskip}{-10pt}
\centering
  \includegraphics[width=3.4in]{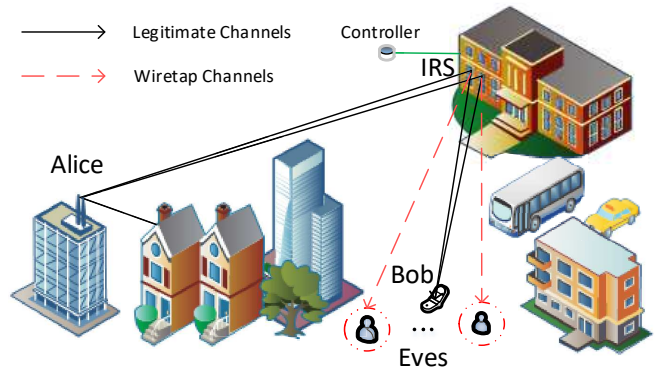}\\
  \caption{System model.}\label{Fig:model}
\end{figure}

Note that all aforementioned works about IRS-aided PLS are based on the assumption
that the channel state information (CSI) of Bob and Eves' channels is perfectly known at the legitimate transmitter. In practice, the CSI acquisition is one of the main challenges for IRS-aided systems due to the lack of RF chains at the IRS. In response this, channel estimation schemes based on element-wise on/off \cite{He2019Cascaded}, near-orthogonal reflection pattern \cite{You2019Intelligent} and compressive sensing \cite{Chen2019Intelligent} have been proposed. However, it is almost impossible to acquire the perfect CSI of Eves' channels even though Eves are assumed to be unscheduled active users in the network. To the best of our knowledge, there is no work considering such problem in IRS-aided mmWave secrecy communication systems. This motivates us to propose a robust BF scheme against the imperfect CSI of Eves' channels. Specifically, assuming that multiple single antenna Eves are stochastically distributed nearby Bob and the angle of arrival (AoA)-based CSI of Eves' channels is imperfectly known at Alice, we formulate robust secure beamforming (RSBF) problems to maximize the worst case ASR against non-colluding and colluding Eves, respectively. Since the formulated problems are non-convex and difficult to solve optimally, we express the wiretap channel based on the weight combination of discrete samples to deal with the imperfect CSI of Eves' channels and then transform the problems into convex forms by applying semidefinite relaxation (SDR) technique. An efficient alternating optimization based RSBF scheme is proposed to solve them sub-optimally. Simulation results show that the proposed RSBF scheme can significantly improve the worst case ASR as compared to the benchmark schemes.

\textsl{Notations}: Throughout our discussions, the distribution of complex Gaussian random variable with mean $\upsilon $ and variance ${\sigma^2}$ is denoted by ${\mathcal{C}\mathcal{N}}\left( \upsilon,\sigma^2 \right)$. ${\left[ x \right]^ + } \!\!=\! \max \left( {0,x} \right)$. Superscript ${{\bf{{A}}}^T}$, ${{\bf{{ A}}}^H}$ and ${{\bf{{A}}}^{ - 1}}$ respectively denote the transpose, the conjugate transpose and the inverse of a matrix ${\bf{{ A}}}$. ${\bf{A}} \underline  \succ     0$ denotes that ${\bf{A}}$ is positive semi-definite. $\otimes$ represents Kronecker product. $eig\left(  \cdot  \right)$ denotes the normalized eigenvector corresponding to the largest eigenvalue of the matrix. ${\left\|\cdot\right\|}$, $tr(\cdot )$ and ${ \mathbb{C}^{M \times N}}$ denote norm, the trace of a matrix and the space of $M \!\times\! N$ complex-valued matrices, respectively. 
\vspace{-1mm}
\section{System Model and Problem Formulation}
\subsection{System Model}
As shown in Fig. 1, we consider a mmWave secrecy communication system, where Alice intends to send confidential information to Bob in the presence of $K$ arbitrarily distributed Eves. Assuming the direct link between Alice and Bob is blocked by obstacles, an IRS managed by a smart controller is deployed to create a line-of-sight (LoS) path to serve Bob. 
Considering that Bob and Eves are all equipped with single antenna, while Alice and IRS are equipped with $M$ antennas and a uniform array with $N$ elements where $N_{az}$ in horizon and $N_{el}$ in vertical, respectively.

The mmWave channels from Alice to IRS, IRS to Bob and the $k$-th Eve are respectively denoted by ${{\bf{H}}_{AR}} $, ${{\bf{h}}_{RB}} $ and ${{\bf{h}}_{{RE}_k}}$, which can be expressed as [1]
\vspace{-1.5mm}
\begin{shrinkeq}{-0.8ex}
\begin{equation}\label{eq2-1}
\begin{small}
{{\bf{H}}_{AR}}\!\! =\!\! \sqrt {1/\!\left( {{\beta _R}{L_R}} \right)} \!\!\sum\limits_{l = 0}^{{L_R} - 1} \!\!{{\alpha _{Rl}}{{\bf{a}}_A}\left( {M,{\phi _l}} \right){\bf{a}}_R^T\left( {N,{\theta _{al}},{\varphi _{al}}} \right)} ,
\end{small}
\end{equation}
\begin{equation}\label{eq2-2}
\begin{small}
{{\bf{h}}_{Ri}} \!\!= \!\!\sqrt {1/\!\left( {{\beta _i}{L_i}} \right)} \sum\limits_{l = 0}^{{L_i} \!-\! 1} {{\alpha _{li}}{{\bf{a}}_R}\left( {N,{\theta _{li}},{\varphi _{li}}} \right)} ,~i \in \left\{ {B,{E_k}} \right\},
\end{small}
\end{equation}
\end{shrinkeq}
where ${\beta _{j}}$ denotes large-scale fading coefficient, $j \in \left\{ {R,B,{E_k}} \right\}$, ${L_{j}}$ is the number of multipaths, $l\!=\!0$ represents the LoS path, ${\alpha}$ denotes small-scale fading that satisfied ${\alpha}\sim {\mathcal{C}\mathcal{N}}\left( 0,1 \right)$, ${{\bf{a}}_A}\left( M, {{\phi}_l} \right){\in \mathbb{C}^{M \times 1}}$ \!and ${{\bf{a}}_R}\left( N, {{\theta},{\varphi} } \right){\in \mathbb{C}^{N \times 1}}$ respectively represent array steering vectors at Alice and IRS, ${\phi}_l$ denotes the angle of departure (AoD) of $l$-th path at Alice, ${{\theta}_{al}}$, ${{\varphi}_{al} }$, ${{\theta}_{li}}$ and ${{\varphi}_{li} }$ denote the AoAs in horizon and vertical, and the AoDs in horizon and vertical of $l$-th path at IRS, respectively. For the $M$-elements array antenna at Alice, the array steering vector can be expressed as
\begin{shrinkeq}{-0.8ex}
\begin{equation}\label{eq2-2}
\begin{small}
{{\bf{a}}}\left(M, {{\phi}_l} \right)\! =\! {\left[ {1,{e^{ - j\frac{{2\pi }}{\lambda }{d_0}\!\cos {\phi }_l}}, \cdots \!,{e^{ - j\frac{{2\pi }}{\lambda }\left( {M - 1} \right){d_0}\!\cos {\phi }_l}}} \right]^T},
\end{small}
\end{equation}
\end{shrinkeq}
where $\lambda$ denotes the mmW wavelength and $d_0$ is the antenna spacing. For the IRS, the array steering vector is ${{\bf{a}}_R}\left( {N, {\theta},{\varphi}} \right) = {{\bf{a}}}\left(N_{az}, {{\theta }} \right) \otimes {{\bf{a}}}\left(N_{el}, {{\varphi }} \right)$.

The received signals at Bob and the $k$-th Eve can be expressed as
\begin{shrinkeq}{-0.8ex}
\begin{equation}\label{eq2-4}
\begin{small}
{y_i} = {{\bf{h}}_{Ri}}{\bf{Q}}{{\bf{H}}_{AR}}{\bf{w}}s + {n},~ i \in \left\{ {B,{E_k}} \right\},
\end{small}
\end{equation}
\end{shrinkeq}
where ${\bf{Q}} { =}  diag\left( \bf{q} \right)$, ${\bf{q}}{ =} \left[{{e^{j{\omega _1}}},{e^{j{\omega _2}}}, \cdots ,{e^{j{\omega _N}}}}\right]$ denotes the diagonal phase-shifting matrix of IRS, ${\omega _n} \in \left[ {0,2\pi } \right)$ represents the phase shift of reflecting signals at IRS by its $n$-th element ($n \!=\! 1,{ \cdots}, N$), ${\bf{w}} {\in \mathbb{C}^{M \times 1}}$ denotes the active BF vector at Alice, $s\!\sim {\mathcal{C}\mathcal{N}}\!\left( {0,1} \right)$ and ${n}\!\sim\! {\mathcal{C}\mathcal{N}}\left( {0,\sigma _0^2} \right)$ denotes the confidential information and complex additive white Gaussian noise (AWGN), respectively. According to (4),  the achievable rate at Bob and the $k$-th Eve can be derived as
\begin{equation}\label{eq2-5}
\setlength\abovedisplayskip{4pt}
\begin{small}
{R_i} = {\log _2}(1 + \frac{{{{\left| {{\bf{q}}{{\bf{H}}_{Ai}}{\bf{w}}} \right|}^2}}}{{\sigma _0^2}}),~i \in \left\{ {B,{E_k}} \right\},
\end{small}
\setlength\belowdisplayskip{4pt}
\end{equation}
where $ {\bf{H}}_{AB}{ = }{{ {diag\left(\! {{{\bf{h}}_{RB}}}\! \right){{\bf{H}}_{AR}}} }}$ and ${{\bf{H}}_{A{E_k}}}{\!=} diag\left(\! {{{\bf{h}}_{R{E_k}}}} \!\right){{\bf{H}}_{AR}}$.

Since multiple Eves are arbitrarily distributed nearby Bob, two scenarios are considered, i.e., colluding and non-colluding eavesdropping, where colluding means Eves are cooperative and non-colluding means Eves are independent. As a result, the ASR can be expressed as
\begin{equation}\label{eq2-6}
\setlength\abovedisplayskip{4pt}
\begin{small}
R_s^t = {\left[ {{R_B} - R_E^t} \right]^ + },~~~t \in \left( {C,I} \right),
\end{small}
\setlength\belowdisplayskip{4pt}
\end{equation}
\vspace{-2mm}where $R_E^C \!=\! {\log _2}\big( {1+\sum\limits_{k = 1}^K {{{{\left| {{\bf{q}}{{\bf{H}}_{A{E_k}}}{\bf{w}}} \right|}^2}} \mathord{\left/
 {\vphantom {{{{\left| {{\bf{q}}{{\bf{H}}_{A{E_k}}}{\bf{w}}} \right|}^2}} {\sigma _{{0}}^2}}} \right.
 \kern-\nulldelimiterspace} {\sigma _{{0}}^2}} } \big)$ and $R_E^I \!\!=\!\! \!\!\!\mathop {\max }\limits_{k \in \left\{ {1, \cdots ,K} \right\}} {\log _2}\big( {1\! +\! {{{{\left| {{\bf{q}}{{\bf{H}}_{A{E_k}}}{\bf{w}}} \right|}^2}} \mathord{\left/
 {\vphantom {{{{\left| {{\bf{q}}{{\bf{H}}_{A{E_k}}}{\bf{w}}} \right|}^2}} {\sigma _{{0}}^2}}} \right.
 \kern-\nulldelimiterspace} {\sigma _{{0}}^2}}} \big)$ denote the ASR for colluding and non-colluding eavesdropping, respectively.
\vspace{-2mm}
\subsection{Problem Formulation}
Based on the various channel acquisition methods discussed in [11]-[13], we assume that the perfect CSI of the legitimate user's channel is known. However, it is almost impossible to obtain the perfect CSI of Eves' channels. The reason is that though they are also active users in the system, they are not currently scheduled. Thus, we consider a more practical and general scenario where the equivalent wiretap channel $ {{\bf{H}}_{A{E_k}}}$ belong to a given range \cite{Lin2018Robust}, i.e.,
\begin{equation}\label{eq2-3}
\begin{small}
\begin{array}{*{20}{l}}
{\Lambda _k} = \left\{ \!{{{\bf{H}}_{A{E_k}}}\!\left| {\left| {{\xi _{{E_k}l}}} \right| \! \in  \!\left[ {\xi _{{E_k}l}^{_{\min}},\xi _{{E_k}l}^{_{\max }}} \right]\!\!, {\theta _{{E_k}l}} \!\in\! \left[ {\theta _{{E_k}l}^{\min},\theta _{{E_k}l}^{\max }} \right]\!\!,} \right.} \right.\\
~~~~~~~~~~~~~~~~~~{\varphi _{{E_k}l}} \!\in \!\left[ {\varphi _{{E_k}l}^{\min }\!,\varphi _{{E_k}l}^{\max }} \right]\left. {,k = 1,2, \cdots  ,K} \right\},
\end{array}
\end{small}
\end{equation}
where ${\xi _{{E_k}l}} {\buildrel \Delta \over =} {{\left| {{\alpha _{{E_k}l}}} \right|} \mathord{\left/
 {\vphantom {{\left| {{\alpha _{{E_k}l}}} \right|} {\sqrt {\left( {{\beta _{{E_k}}}{L_i}} \right)} }}} \right.
 \kern-\nulldelimiterspace} {\sqrt {\left( {{\beta _{{E_k}}}{L_i}} \right)} }}$ denotes the amplitude of $l$-th path from Alice to $E_k$, superscripts $min$ and $max$ respectively denote the lower and upper bound, which are available based on historical CSI as we assume that Eves (unscheduled users in the network) have been served at the previous time slots.

Due to the imperfect CSI of Eves' channels, we aim to jointly design the active BF vector $\bf{w}$ and passive BF vector $\bf{q}$ to maximize the worst case ASR, subject to the total transmit power constraint at Alice and unit-modulus constraints at IRS. Accordingly, the problem can be formulated as
\begin{equation}\label{eq2-7}
\begin{small}
\begin{array}{l}
{\rm{P0:}}~\mathop {\max }\limits_{{\bf{w}},{\bf{q}}} \mathop {{\rm{min}}}\limits_{~{{\bf{H}}_{A{E_k}}} } {{R}}_s^t,\\
~~~~~~~{\rm{s.t.}}~~~C1:~{\left\| {\bf{w}} \right\|^2} \le {P_{\max }},\\
~~~~~~~~~~~~~~C2:~\left| {{q_n}} \right| = 1,~~\forall~ n=1, \cdots , N,
\end{array}
\end{small}
\end{equation}
where ${q_n} \! =\! {e^{j{w_n}}}$. Considering the colluding and non-colluding eavesdropping scenarios, the optimization problem P0 can be reformulated as the following two problems,
\vspace{-2mm}
\begin{equation}\label{eq2-8}
\begin{small}
\begin{array}{l}
{\rm{P1:}}~\mathop {\max }\limits_{{\bf{w}},{\bf{q}}} {\mathop {{\!{\min}}}\limits_{~{{\bf{H}}_{A{E_k}}} }} ~{\log _2}\left( {\frac{{{{\left| {{\bf{q}}{{\bf{H}}_{AB}}{\bf{w}}} \right|}^2} + \sigma _0^2}}{{\sum\limits_{k = 1}^K {{{\left| {{\bf{q}}{{\bf{H}}_{A{E_k}}}{\bf{w}}} \right|}^2}}  + \sigma _{{0}}^2}}} \right),\\
~~~~~~~{\rm{s.t.}}~~C1,~C2,\\
\end{array}
\end{small}
\end{equation}
\vspace{-2mm}
\begin{equation}\label{eq2-9}
\begin{small}
\begin{array}{l}
~~~~~{\rm{P2:}}~\mathop {\max }\limits_{{\bf{w}},{\bf{q}}} \mathop {{{\min}}}\limits_k \!\mathop {{\!{\min}}}\limits_{~{{\bf{H}}_{A{E_k}}} } {\log _2}\left( {\frac{{{{\left| {{\bf{q}}{{\bf{H}}_{AB}}{\bf{w}}} \right|}^2} + \sigma _0^2}}{{{{{\left| {{\bf{q}}{{\bf{H}}_{A{E_k}}}{\bf{w}}} \right|}^2}} + \sigma _{0}^2}}} \right),\\
~~~~~~~~~~~~{\rm{s.t.}}~~C1,~C2,\\
\end{array}
\end{small}
\end{equation}
Obviously, it's difficult to solve such optimization problems optimally because of non-convex objective functions, coupled variables and non-convex unit-modulus constraints. Furthermore, the imperfect CSI of Eves' channels makes them more challenging. 
In the following sections, we propose an efficient robust BF algorithm to approximately solve them.

\section{Robust Beamforming Algorithm for P1}
First, to deal with the imperfect CSI of Eves' channels, we construct a convex hull of ${\Lambda}_k$ based on weighted sum of $D_K$ discrete samples \cite{Boyd2004Convex}, which is written as
\begin{equation}\label{eq3-1}
\begin{small}
{{\bf{\Xi}}  _k} \!=\! \left\{ {\sum\limits_{t = 1}^{{D_K}} {{\mu _{k,t}}{\bf{G}}_{k,t}^H{{\bf{G}}_{k,t}}\left| {\sum\limits_{t = 1}^{{D_K}} {{\mu _{k,t}} \!=\! 1,{\mu _{k,t}} \ge 0} } \right.} } \right\},\forall k,
\end{small}
\end{equation}
where ${{\bf{G}}_{k,t}} { = }{\left( {diag\left({\bf{h}}_{R{E_k}}\right){\bf{H}}_{AR}} \right)_{t}}$ and ${{\mu _{k,{t}}}}$ is the weighted coefficient of the $t$-th discrete sample. As the \!${\log _2}\!\left(  \cdot  \right)$\! is a monotonically increasing function, P1 can be equivalently written as
\begin{equation}\label{eq3-1}
\begin{small}
\begin{array}{*{20}{l}}
{\rm{P1.1:}}~\mathop {\max }\limits_{{\bf{w}},{\bf{q}}} \mathop {{\rm{min}}}\limits_{\bf{{\Xi} _k}} \frac{{{{\bf{w}}^H}{\bf{H}}_{AB}^H{{\bf{q}}^H}{\bf{q}}{{\bf{H}}_{AB}}{\bf{w}} + \sigma _0^2}}{{\sum\limits_{k = 1}^K {\sum\limits_{t = 1}^{{D_K}} {{\mu _{k,t}}} {{\bf{w}}^H}{\bf{G}}_{k,t}^H{{\bf{q}}^H}{\bf{q}}{{\bf{G}}_{k,t}}{\bf{w}} + \sigma _{{0}}^2} }},\\
~~~~~~~~~{\rm{s.t.}}~~C1,~C2.\\
\end{array}
\end{small}
\end{equation}
Note that the optimization problem P1.1 is still intractable. However, it is observed that the worst case of objective function can be efficiently solved if $\bf{w}$ and $\bf{q}$ are fixed, and on the other hand, if the worst case of wiretap channel is given, $\bf{w}$ and $\bf{q}$  can be iteratively optimized. This inspires us to sub-optimally solve P1.1 by two-layer loop iteration as detailed in following.
\vspace{-3mm}
\subsection{Optimizing ${\mu _{k,t}}$ for Given $\bf{w}$ and $\bf{q}$}
For given $\bf{w}$ and $\bf{q}$, P1.1 can be transformed as
\begin{equation}\label{eq3-1}
\begin{small}
\begin{array}{*{20}{l}}
{\rm{P1.2:}}~\mathop {{\rm{min}}}\limits_{\left\{ {{\mu _{k,t}}} \right\}} \frac{{{{\bf{w}}^H}{\bf{H}}_{AB}^H{{\bf{q}}^H}{\bf{qH}}_{AB}{\bf{w}} + \sigma _0^2}}{{\sum\limits_{k = 1}^K {\sum\limits_{t = 1}^{{D_K}} {{\mu _{k,t}}{{\bf{w}}^H}{\bf{G}}_{k,t}^H{{\bf{q}}^H}{\bf{q}}{{\bf{G}}_{k,t}}{\bf{w}}}  + \sigma _{{0}}^2} }}.
\end{array}
\end{small}
\end{equation}
Let ${\bf{\bar w}} {=} {\bf{q}}{{\bf{G}}_{k,t}}{\bf{w}}$, then the optimal ${\left\{ {{\mu _{k,t}}} \right\}}$ is obtained by maximizing $\sum\limits_{k = 1}^K {\sum\limits_{t = 1}^{{D_K}} {\mu _{k,t}}{{{{\bf{\bar w}}}^H}} {\bf{\bar w}}}$. According to the Cauchy-Schwarz's inequality, we have
\begin{equation}\label{eq3-2}
\begin{small}
{\left( {\sum\limits_{t = 1}^{{D_K}} {\mu _{k,t}}{{{{\bf{\bar w}}}^H}} {\bf{\bar w}}} \right)^2} \le \left( {\sum\limits_{t = 1}^{{D_K}} {\mu _{k,t}^2} } \right)\sum\limits_{t = 1}^{{D_K}} {{{\left( {{{{\bf{\bar w}}}^H}{\bf{\bar w}}} \right)}^2}} .
\end{small}
\end{equation}
The inequality (14) holds true only when ${\mu _{k,1}}/{{{\bf{\bar w}}}^H}{\bf{\bar w}} = {\mu _{k,2}}/{{{\bf{\bar w}}}^H}{\bf{\bar w}} =  \cdots  = {\mu _{k,{D_K}}}/{{{\bf{\bar w}}}^H}{\bf{\bar w}}$ is satisfied. Based on the constraint $\sum\limits_{t = 1}^{{D_K}} {{\mu _{k,t}}} = 1$, we can compute ${{\mu _{k,t}}}$ as [14]
\vspace{-2mm}
\begin{equation}\label{eq3-3}
\begin{small}
{\mu _{k,t}} = {{{\bf{\bar w}}}^H}{\bf{w}}{\left( {\sum\limits_{t = 1}^{{D_K}} {{{\bf{w}}^H}} {\bf{w}}} \right)^{ - 1}}.
\end{small}
\end{equation}
\vspace{-8mm}
\subsection{Optimizing $\bf{w}$ and $\bf{q}$ for Given ${\mu _{k,t}}$}
For given ${\mu _{k,t}}$, P1.1 can be transformed as following
\begin{equation}\label{eq3-4}
\begin{small}
\begin{array}{l}
{\rm{P1.3:}}~\mathop {\max }\limits_{{\bf{w}},{\bf{q}}} \frac{{{{\bf{w}}^H}{\bf{H}}_{AB}^H{{\bf{q}}^H}{\bf{q}}{{\bf{H}}_{AB}}{\bf{w}} + \sigma _0^2}}{{\sum\limits_{k = 1}^K {\sum\limits_{t = 1}^{{D_K}} {{\mu _{k,t}}{{\bf{w}}^H}{\bf{G}}_{k,t}^H{{\bf{q}}^H}{\bf{q}}{{\bf{G}}_{k,t}}} } {\bf{w}} + \sigma _{{0}}^2}},\\
~~~~~~~~~{\rm{s.t.}}~~C1,~C2.\\
\end{array}
\end{small}
\end{equation}
As $\bf{w}$ and $\bf{q}$ are coupled, we again apply alternating optimization to approximately solve them \cite{Wu2019Intelligent}.

1) Optimizing $\bf{w}$ for given $\bf{q}$ and ${\mu _{k,t}}$. For given $\bf{q}$ and ${\mu _{k,t}}$, P1.3 can be rewritten as
\begin{equation}\label{eq3-4}
\begin{small}
\begin{array}{l}
{\rm{P1.4:}}~\mathop {\max }\limits_{\bf{w}} \frac{{{{\bf{w}}^H}\left( {{\bf{H}}_{AB}^H{{\bf{q}}^H}{\bf{q}}{{\bf{H}}_{AB}}} \right){\bf{w}} + \sigma _0^2}}{{{{\bf{w}}^H}\left( {\sum\limits_{k = 1}^K {\sum\limits_{t = 1}^{{D_K}} {{\mu _{k,t}}{\bf{G}}_{k,t}^H{{\bf{q}}^H}{\bf{q}}{{\bf{G}}_{k,t}}} } } \right){\bf{w}} + \sigma _{{0}}^2}},\\
~~~~~~~~~{\rm{s.t.}}~~C1,\\
\end{array}
\end{small}
\end{equation}
whose optimal solution is \cite{Cui2019Secure}
\begin{equation}\label{eq3-5}
\begin{small}
{{\bf{w}}_{opt}} \!\!=\!\! \sqrt {{P_{\max }}} eig\left( {{{\left( {{{\bf{A}}_1} + {\tau} {{\bf{I}}_M}} \right)}^{ - 1}}\!\!\left( {{{\bf{B}}_1} + {\tau}{{\bf{I}}_M}} \right)} \right),
\end{small}
\end{equation}
where \!${{\bf{A}}_1} { = }\sum\limits_{k = 1}^K  \sum\limits_{t = 1}^{{D_K}}  {\mu _{k,t}}{\bf{G}}_{k,t}^H{{\bf{q}}^H}{\bf{qG}}_{k,t}$, \!${{\bf{B}}_1} { =} {\bf{H}}_{AB}^H{{\bf{q}}^H}{\bf{q}}{{\bf{H}}_{AB}}$,
${\tau} { =} {{\sigma _{{0}}^2}\mathord{\left/
 {\vphantom {{\sigma _{{0}}^2} {{P_{\max }}}}} \right.
 \kern-\nulldelimiterspace} {{P_{\max }}}}$ and ${\bf{I}}_M $ is an $M \times M$ identity matrix.

\vspace{3mm}
2) Optimizing $\bf{q}$ for given $\bf{w}$ and ${\mu _{k,t}}$. Similarly, given $\bf{w}$ and ${\mu _{k,t}}$, P1.3 can be reformulated as
\begin{shrinkeq}{-0.5ex}
\begin{equation}\label{eq3-6}
\begin{small}
\begin{array}{l}
{\rm{P1.5:}}~\mathop {\max }\limits_{\bf{q}} \frac{{{\bf{q}}\left( {{{\bf{H}}_{AB}}{\bf{w}}{{\bf{w}}^H}{\bf{H}}_{AB}^H} \right){{\bf{q}}^H} + \sigma _0^2}}{{{\bf{q}}\left( {\sum\limits_{k = 1}^K {\sum\limits_{t = 1}^{{D_K}} {{\mu _{k,t}}{{\bf{G}}_{k,t}}{\bf{w}}{{\bf{w}}^H}{\bf{G}}_{k,t}^H} } } \right){{\bf{q}}^H} + \sigma _0^2}},\\
~~~~~~~~~{\rm{s.t.}}~~C2.\\
\end{array}
\end{small}
\end{equation}
\end{shrinkeq}
Let ${{\bf{A}}_2} {=} {{\bf{H}}_{AB}}{\bf{w}}{{\bf{w}}^H}{\bf{H}}_{AB}^H$, ${{\bf{B}}_2} {=} \sum\limits_{k = 1}^K {\sum\limits_{t = 1}^{{D_K}} {{\mu _{k,t}}{{\bf{G}}_{k,t}}{\bf{w}}{{\bf{w}}^H}{\bf{G}}_{k,t}^H} } $ and ${\bf{Q}}_1 { =}{{\bf{q}}^H}{\bf{q}}$, then P1.5 can be rewritten  as
\begin{equation}\label{eq3-7}
\begin{small}
\begin{array}{l}
{\rm{P1.6:}}~\mathop {\max }\limits_{{\bf{Q}}_1 \underline  \succ  0} ~\frac{{tr\left( {{\bf{A}}_2}{{\bf{Q}}_1} \right) + \sigma _0^2}}{{tr\left( {{\bf{B}}_2}{{\bf{Q}}_1} \right) + \sigma _{{0}}^2}},\\
~~~~~~~~~{\rm{s.t.}}~~tr\left( {{{\bf{E}}_n}{{\bf{Q}}_1}} \right) = 1,~~\forall~ n=1, \cdots, N,\\
~~~~~~~~~~~~~~~{\rm{rank}}\left( {{\bf{Q}}_1} \right) = 1,
\end{array}
\end{small}
\end{equation}
where ${\left[ {{{\bf{E}}_n}} \right]_{a,b}}$ denotes the $\left( {a,b} \right)$-th element of ${{{\bf{E}}_n}}$ and satisfies that ${\left[ {{{\bf{E}}_n}} \right]_{a,b}}\!=\!1$ only when $n\!=a\!=b$, otherwise, ${\left[ {{{\bf{E}}_n}} \right]_{a,b}}=0$. Note P1.6 is still mathematically intractable as the constraint ${\rm{rank}}\left( {{\bf{Q}}_1} \right) = 1$. By defining $\varsigma {{ =}}{1 \mathord{\left/
 {\vphantom {1 {\left( {tr\left( {{\bf{B}}_2}{{\bf{Q}}_1} \right) + \sigma _{{0}}^2} \right)}}} \right.
 \kern-\nulldelimiterspace} {\left( {tr\left( {{\bf{0}}_2}{{\bf{Q}}_1} \right) \!+\! \sigma _{{E}}^2} \right)}}$, ${{\bf{Q}}_2} \!\!=\!\! \varsigma {{\bf{Q}}_1}$ and applying SDR, we can transform P1.6 as a convex semidefinite programming (SDP) problem, i.e.,
 \begin{equation}\label{eq3-8}
\begin{small}
 \begin{array}{l}
~~{\rm{P1.7:}}~\mathop {\max }\limits_{{\bf{Q}}_2 \underline\succ 0,\varsigma  \ge 0} ~tr\left( {{\bf{A}}_2} {{\bf{Q}}_2}\right) + \varsigma \sigma_0^2 ,\\
~~~~~~~~~~~~~~{\rm{s.t.}}~~tr\left( {{\bf{B}}_2} {{\bf{Q}}_2} \right) + \varsigma \sigma_{0}^2 = 1,\\
~~~~~~~~~~~~~~~~~~~~tr\left( {{{\bf{E}}_n}{{\bf{Q}}_2}} \right) = \varsigma ,~~\forall~ n=1, \cdots, N.
\end{array}
\end{small}
\end{equation}
As to P1.7, it can be optimally solved by using convex optimization toolbox, e.g. CVX. Note the constraint ${\rm{rank}}\left( {{{\bf{Q}}_1}} \right)\!\! =\!\! 1$ may be not always guaranteed. For the case ${\rm{rank}}\left( {{\bf{Q}}_1} \right) \!\!\ne\!\! 1$, similar to that in \cite{Wu2019Intelligent}, Gaussian randomization method can be used to recover $\bf{q}$ but omitted here for brevity.

The proposed two-layer iterative RSBF algorithm to solve P1 is summarized as {\bf{Algorithm 1}} where $\varepsilon$ denotes a small threshold. The main complexity of Algorithm 1 is due to the Step 6, 9 and 10. Specifically, the complexity of Step 6 is $ {\mathcal{O}}\left({{D_K}KMN }\right)$, whilst it is $ {\mathcal{O}}\left({N^3}\right)$ in Step 9 and ${\mathcal{O}}\big(\left( {{N}\! +\! 1} \right)^{3.5} \big)$ in Step 10. Thus, the overall complexity of Algorithm 1 is $ {\mathcal{O}}\big( {{I_{out }}{I_{in }}\big({{D_K}KMN \!+\! {N^3}\! +\! \left( {{N}\! +\! 1} \right)^{3.5}} \big)} \big)$, where ${I_{out }}$ and ${I_{in }}$ denote the outer and inner iteration numbers required for convergence, respectively. 

\begin{algorithm}[ht]
\caption{Proposed RSBF Algorithm for P1}
\begin{algorithmic}[1]
\STATE{{\bf{Input}}: ${P_{max}, diag\left({{\bf{h}}_{RD}}\right) {{\bf{H}}_{AR}} }, L, K, D_K, \varepsilon$.}
\STATE{ Initialize ${{\bf{w}}^{\left( 0 \right)}}\!\!\! =\!\! {{{{\bf{H}}_{AB}}\left( {1,:} \right)} \mathord{\left/
 {\vphantom {{{{\bf{H}}_{AB}}\left( {1,:} \right)} {\left\| {{{\bf{H}}_{AB}}\left( {1,:} \right)} \right\|}}} \right.
 \kern-\nulldelimiterspace} {\left\| {{{\bf{H}}_{AB}}\left( {1,:} \right)} \right\|}}, {\bf{Q}}\!\! =\!\! {{\bf{I}}_N}, g\!\!=\!\!0, m\!=\!0, R_{s - in}^{\left( 0 \right)}=0, R_{s - out}^{\left( 0 \right)}=0 $}.
\STATE{Let ${\theta _{{E_k}l}}\! \sim \!U\left[ {\theta _{{E_k}l}^{_{\min }},\theta _{{E_k}l}^{_{\max }}} \right]$, ${\varphi _{{E_k}l}} \!\sim \!U\left[ {\varphi _{{E_k}l}^{_{\min }},\varphi _{{E_k}l}^{_{\max }}} \right]$ and ${\xi _{{E_k}l}}\! = \!\xi _{{E_k}l}^{_{\min }}$, construct ${{{\bf{G}}_{k,t}}}$, where ${k = 1,2, \cdots ,K}; t = 1,2, \cdots ,{D_K}; l = 1, \cdots ,L$. }
\STATE{{~\bf{repeat}}}
\STATE{~g:=g+1;}
\STATE{~Update $\mu _{k,t}^{\left( g \right)}$ based on (14) for given ${{\bf{w}}^{\left( {g - 1} \right)}}$ and ${{\bf{q}}^{\left( {g - 1} \right)}}$.}
\STATE{~~{\bf{repeat}}}
\STATE{~~m:=m+1;}
\STATE{~~Calculate ${\bf{w}}_{opt}^{\left( {m} \right)}$ for given $\mu _{k,t}^{\left( g \right)}$ and ${{\bf{q}}^{\left( {m - 1} \right)}}$} as (17).
\STATE{~~Compute ${{\bf{q}}^{\left( {m } \right)}}$\! for given\! $\mu _{k,t}^{\left( g \right)}$ and ${\bf{w}}_{opt}^{\left( {m} \right)}$} \!by solving (20).
\STATE{~~Set $R_{s - in}^{\left( m \right)} = f\left( {\mu _{k,t}^{\left( g \right)},{\bf{w}}_{opt}^{\left( m \right)},{{\bf{q}}^{\left( m \right)}}} \right)$;}
\STATE{~~{\bf{until}}~$\left| {R_{s - in}^{\left( m \right)} - R_{s - in}^{\left( {m - 1} \right)}} \right| \le \varepsilon $}.
\STATE{~Set ${\bf{w}}_{opt}^{\left( g \right)} = {\bf{w}}_{opt}^{\left( m \right)}$ and ${{\bf{q}}^{\left( g \right)}} = {{\bf{q}}^{\left( m \right)}}$;}
\STATE{~Set $R_{s - out}^{\left( g \right)} = f\left( {\mu _{k,t}^{\left( g \right)},{\bf{w}}_{opt}^{\left( g \right)},{{\bf{q}}^{\left( g \right)}}} \right)$;}
\STATE{~{\bf{until}}~$\left| {R_{s - out}^{\left( g \right)} - R_{s - out}^{\left( {g - 1} \right)}} \right| \le \varepsilon $}.
\STATE{{\bf{Output}}: ${\bf{w}}_{opt}, {\bf{q}}$}
\end{algorithmic}
\end{algorithm}

\section{Robust Beamforming Algorithm for P2}
For the scenario that Eves are non-colluding, by omitting ${\log _2}\left(  \cdot  \right)$ and some trivial calculations, P2 can be rewritten as
\begin{equation}\label{eq4-1}
\begin{small}
\begin{array}{*{20}{l}}
{\rm{P2.1:}}~\mathop {\min }\limits_{{\bf{w}},{\bf{q}}} \mathop {{\rm{max}}}\limits_k \mathop {{\rm{max}}}\limits_{{{\bf{G}}_k}} \frac{{{{\bf{w}}^H}{\bf{G}}_k^H{{\bf{q}}^H}{\bf{qG}}_k^{}{\bf{w}} + \sigma _{{0}}^2}}{{{{\bf{w}}^H}{\bf{H}}_{AB}^H{{\bf{q}}^H}{\bf{q}}{{\bf{H}}_{AB}}{\bf{w}} + \sigma _0^2}},\\
~~~~~~~~~{\rm{s.t.}}~~C1,~C2,\\
\end{array}
\end{small}
\end{equation}
where ${{\bf{G}}_{k}} { = }{ {diag\left({\bf{h}}_{R{E_k}}\right){\bf{H}}_{AR}} }$.  Note that a similar approach as proposed in Section ${\rm I}{\rm I}{\rm I}$ can be adopted to handle the imperfect CSI of Eves' channels, and $\mu _{k,t}$ can be updated according to (15). Therefore, P2.1 can be reduced as
\begin{equation}\label{eq4-10}
\begin{small}
\begin{array}{*{20}{l}}
{{\rm{P2.2:}}~\mathop {\min }\limits_{{\bf{w}},{\bf{q}}} \mathop {{\rm{max}}}\limits_k \frac{{\sum\limits_{t = 1}^{{D_K}} {\mu _{k,t}^{}} {{\bf{w}}^H}{\bf{G}}_{k,t}^H{{\bf{q}}^H}{\bf{qG}}_{k,t}^{}{\bf{w}} + \sigma _{{0}}^2}}{{{{\bf{w}}^H}{\bf{H}}_{AB}^H{{\bf{q}}^H}{\bf{q}}{{\bf{H}}_{AB}}{\bf{w}} + \sigma _0^2}}},\\
~~~~~~~~~{\rm{s.t.}}~~C1,~C2\\
\end{array}
\end{small}
\end{equation}
Similar to P1.1, P2.2 can be approximately solved by iteratively optimizing $\bf{q}$ and $\bf{w}$.
\vspace{-0.2mm}
\subsection{Optimizing $\bf{w}$ for Given $\bf{q}$ and ${ {{\mu _{k,t}}} }$}
For given $\bf{q}$ and ${ {{\mu _{k,t}}}}$, after some mathematical operation, P2.2 can be rewritten as
\begin{shrinkeq}{-0.8ex}
\begin{equation}\label{eq4-7}
\begin{small}
\begin{array}{*{20}{l}}
{\rm{P2.3:}}~{\mathop {\min }\limits_{{\bf{W}} \underline\succ 0} \mathop {{\rm{max}}}\limits_k \frac{{tr\left( {{ {{{\bf{C}}_k}} }{\bf{W}}} \right) + \sigma _{{0}}^2}}{{tr\left(  {{\bf{A}}_3}{{\bf{W}}} \right) + \sigma _0^2}}},\\
{~~~~~~~~~{\rm{s.t.}}~~tr\left( {\bf{W}} \right) \le {P_{\max }}},\\
~~~~~~~~~~~~~~~{{\rm{rank}}\left( {\bf{W}} \right) = 1},
\end{array}
\end{small}
\end{equation}
\end{shrinkeq}
\vspace{-0.5mm}where ${{\bf{C}}_k}\! =\! \sum\limits_{t = 1}^{{D_K}}  {\mu _{k,t}}{\bf{G}}_{k,t}^H{{\bf{q}}^H}{\bf{qG}}_{k,t}$, ${{\bf{A}}_3}\! =\! {\bf{H}}_{AB}^H{{\bf{q}}^H}{\bf{q}}{{\bf{H}}_{AB}}$ and ${\bf{W}} {=}{\bf{w}}{{\bf{w}}^H}$.
By introducing auxiliary variables ${r}$, $v$ and ${\bf{X}}\!=\!v{{\bf{W}}}$, P2.3 can be converted into
\begin{equation}\label{eq4-8}
\begin{small}
\begin{array}{l}
{\rm{P2.4:}}~\mathop {\min }\limits_{{\bf{X}} \underline\succ 0,v \ge 0,r} ~~r,\\
~~~~~~~~~~~~~{\rm{s.t.}}~~tr\left( {{ {{{\bf{C}}_k}}}{\bf{X}}} \right) + v\sigma _{{0}}^2 \le r,\forall k,\\
~~~~~~~~~~~~~~~~~~~tr\left( {{\bf{A}}_3}{{\bf{X}}} \right) + \sigma _0^2 \ge 1,\\
~~~~~~~~~~~~~~~~~~~tr\left( {\bf{X}} \right) \le v{P_{\max }},\\
~~~~~~~~~~~~~~~~~~~{\rm{rank}}\left( {\bf{X}} \right) = 1.
\end{array}
\end{small}
\end{equation}
By applying SDR, the approximate solution of $\bf{X}$ and $v$ are obtained, and the approximate solution of $\bf{W}$ can be computed as ${\bf{W}}\! =\! {{\bf{X}} \mathord{\left/
 {\vphantom {{\bf{X}} v}} \right.
 \kern-\nulldelimiterspace} v}$. Finally, based on the singular value decomposition (SVD) of $\bf{W}$, $\bf{w}$ can be recovered as the eigenvector of maximal eigenvalue, if $\rm{rank}\left( {\bf{W}} \right) \!=\! 1$ satisfied. Otherwise, Gaussian randomization method [4] can be used to recover $\bf{w}$.
\vspace{-4mm}
\subsection{Optimizing $\bf{q}$ for Given $\bf{w}$ and ${ {{\mu _{k,t}}} }$}
For given $\bf{w}$ and and ${ {{\mu _{k,t}}} }$, P2.2 can be simplified as
\begin{shrinkeq}{-0.6ex}
\begin{equation}\label{eq4-2}
\begin{small}
\begin{array}{*{20}{l}}
{\rm{P2.5:}}~\mathop {\min }\limits_{{\bf{Q}}_1 \underline\succ 0} \mathop {{\rm{max}}}\limits_k \frac{{tr\left( {{{\bf{F}}_k}{{\bf{Q}}_1}} \right) + \sigma _{{0}}^2}}{{tr\left( {{\bf{A}}_2} {{\bf{Q}}_1}\right) + \sigma _0^2}},\\
~~~~~~~~~~{\rm{s.t.}}~~tr\left( {{{\bf{E}}_n}{{{\bf{Q}}_1}}} \right) = 1,~~\forall n,\\
~~~~~~~~~~~~~~~~{\rm{rank}}\left( {{{\bf{Q}}_1}} \right) = 1,
\end{array}
\end{small}
\end{equation}
\end{shrinkeq}
where ${{\bf{F}}_k} = \sum\limits_{t = 1}^{{D_K}}  {\mu _{k,t}}{{\bf{G}}_{k,t}}{\bf{w}}{{\bf{w}}^H}{\bf{G}}_{k,t}^H$.
By introducing auxiliary variables $r'$, $v'$ and ${\bf{S}}=v'{{\bf{Q}}_1}$, we can convert P2.5 into the following problem
\begin{shrinkeq}{-0.6ex}
\begin{equation}\label{eq4-3}
\begin{small}
\begin{array}{l}
{\rm{P2.6:}}~\mathop {\min }\limits_{{{\bf{S}}} \underline\succ 0,v' \ge 0,r'} ~~r',\\
~~~~~~~~~~~~~{\rm{s.t.}}~~tr\left( {{{\bf{F}}_k}} {\bf{S}}\right) + v'\sigma _{{0}}^2 \le r',~~\forall k,\\
~~~~~~~~~~~~~~~~~~~tr\left( {{{\bf{A}}_2}{\bf{S}}} \right) + \sigma _0^2 \ge 1,\\
~~~~~~~~~~~~~~~~~~~tr\left( {{{\bf{E}}_n}{\bf{S}}} \right) = 1,~~\forall n,\\
~~~~~~~~~~~~~~~~~~~{\rm{rank}}\left( {\bf{S}} \right) = 1.
\end{array}
\end{small}
\end{equation}
\end{shrinkeq}
Similarly, by applying SDR, P2.6 can be converted into a convex problem and thus effectively solved. For the case that ${\rm{rank}}\big( {{{\bf{Q}}_1}} \big) \ne  1$, Gaussian randomization method is again used to recover $\bf{q}$.

Note that the overall algorithm to solve P2 is similar to {\bf{Algorithm 1}} and thus omitted for simplicity. The main difference is that ${\bf{w}}_{opt}^{\left( {m} \right)}$ in Step 9 and ${{\bf{q}}^{\left( {m } \right)}}$ in Step 10 are calculated by solving (25) and (27), respectively. 
\vspace{0mm}
\section{Simulation results}
\vspace{-1mm}
\begin{figure}
\centering
  \includegraphics[width=3.4in]{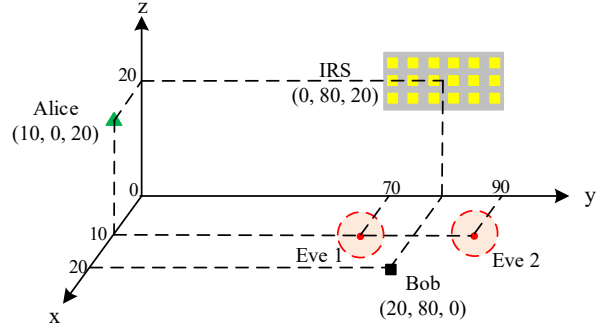}\\
  \caption{Simulation setup.}\label{Fig:model}
\end{figure}
In this section, numerical results are provided to demonstrate the effectiveness of the proposed RSBF schemes. As shown in Fig. 2, it is assumed that Alice, IRS and Bob are respectively located at (10, 0, 20), (0, 80, 20) and (20, 80, 0) in meters. While two Eves (i.e., $K=2$) are located within two circular regions, nearby (10, 70, 0) and (10, 90, 0) in meters, respectively. The  large-scale fading ${\beta _{j}}$ is taken as ${\beta _j}\left( {c,d} \right) = {\varsigma _0} - 10c{\log _{10}}\left( {{d_j}} \right)$  where $d_j$ denotes the signals propagation distance and the  the path loss exponents of LoS and No-LoS are set as ${c_{LoS}} \!=\! 2$ and ${c_{NLoS}}\! =\! 5$, respectively. We set $M\!=\!16$, $N\!=\!N_{az} \times N_{el}\!=\!4 \times 4 $, $\varepsilon \! =\! {10^{ - 3}}$, $L\!=\!4$, ${\varsigma _0}\! =\!  - 61.4~dB$ and $\sigma _0^2\!= \!-110~dBm$. The maximum ratio transmission (MRT)-based scheme, perfect CSI based scheme \cite{Cui2019Secure} and average based scheme are adopted for performance comparison, where MRT-based scheme means that the passive BF vector is optimized to maximize $\left\| {{{\bf{h}}_{RB}}{\bf{Q}}{{\bf{H}}_{AR}}} \right\|$ and the active BF vector is designed based on MRT, while the average-based scheme means that the active and passive BF vectors are jointly optimized based on average of Eves' AoAs.

\begin{figure}
\centering
  \includegraphics[width=3.0in]{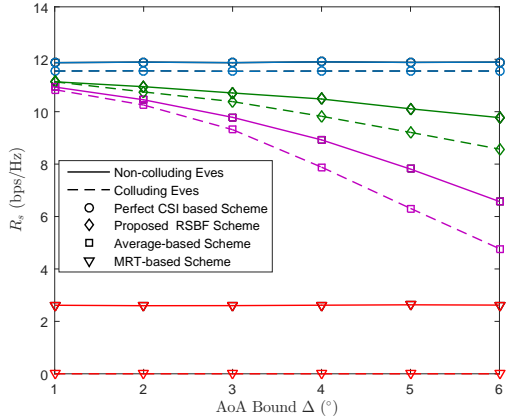}\\
  \caption{ASR versus AoA error bound, $P_{max}=1~W$.}\label{Fig:model}
\end{figure}

Fig.3 depicts the ASR versus the AoA-based uncertainty of Eves for different BF algorithms, where the perfect CSI based scheme is provided as the upper bound. 
It can be observed that the ASR of the MRT-based scheme keeps stable as the error bound of AoA increases but is worse than others. This is due to the fact that the ASR of MRT-based scheme only depends on the CSI of legitimate channels. Meanwhile, the ASR of the proposed RSBF scheme decreases slowly as the error bound of AoA increases. In contrast, the ASR of the average-based scheme significantly decreases, which validates the robustness of proposed RSBF scheme.

Fig.4 plots the ASR versus the transmission power for different BF algorithms with both colluding and non-colluding eavesdropping scenarios. It can be observed that the gap between upper bound and the proposed RSBF scheme is smaller than that between upper bound and the MRT-based scheme, which demonstrates the effectiveness and superiority of the proposed RSBF scheme. Meanwhile, the ASR of perfect CSI scheme and proposed RSBF scheme significantly increase, but that of MRT-based scheme increases slowly and even keeps zero for the colluding eavesdropping scenario as the transmission power increases. This is due to the fact that the MRT-based scheme aims to maximize the achievable rate of Bob while ignoring Eves, which results in significant information leakage since the Eves are located nearby Bob, while the proposed RSBF scheme can effectively prevent eavesdropping against the imperfect CSI of Eves' channels.

\begin{figure}
\centering
  \includegraphics[width=3.0in]{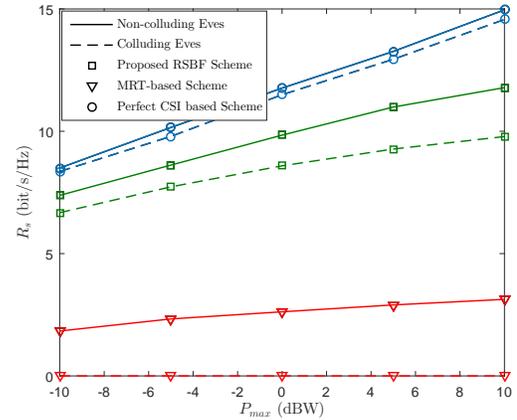}\\
  \caption{ASR versus transmission power, $\Delta =5^ \circ $.}\label{Fig:model}
\end{figure}
\vspace{-1.5mm}
\section{Conclusion}\label{conclusion}
\vspace{-1mm}
In this letter, we have investigated an RSBF scheme in an IRS-aided mmWave secrecy system, against multiple single antenna Eves. Considering the AoA-based CSI of Eves' channels is imperfectly known by the legitimate transmitter, we formulate optimization problems to maximize the worst case of ASR for both colluding and non-colluding eavesdropping scenarios, and propose a robust algorithm to solve such non-convex problems. Numerical results are provided to illustrate the effectiveness of proposed RSBF schemes.

\end{document}